# Open Problems in Engineering and Quality Assurance of Safety Critical Machine Learning Systems

Hiroshi Kuwajima, Hirotoshi Yasuoka, Toshihiro Nakae
DENSO CORPORATION
Kariya, Aichi
{HIROSHI_KUWAJIMA,HIROTOSHI_YASUOKA,TOSHIHIRO_NAKAE}@denso.co.jp

## ABSTRACT

Fatal accidents are a major issue hindering the wide acceptance of safety-critical systems using machine-learning and deep-learning models, such as automated-driving vehicles. Quality assurance frameworks are required for such machine-learning systems, but there are no widely accepted and established quality-assurance concepts and techniques. At the same time, open problems and the relevant technical fields are not organized. To establish standard quality assurance frameworks, it is necessary to visualize and organize these open problems in an interdisciplinary way, so that the experts from many different technical fields may discuss these problems in depth and develop solutions. In the present study, we identify, classify, and explore the open problems in quality assurance of safety-critical machine-learning systems, and their relevant corresponding industry and technological trends, using automated-driving vehicles as an example. Our results show that addressing these open problems requires incorporating knowledge from several different technological and industrial fields, including the automobile industry, statistics, software engineering, and machine learning.

## 1 INTRODUCTION

The recent development in the machine learning techniques such as deep neural networks has led to the widespread application of systems that assign advanced environmental perception and decision-making to computer logics learned from big data, instead of manually built rule-based logics [4]. Available big data and affordable high performance computing such as deep learning framework on off-the-shelf GPU [24] made highly complex machine learning techniques practical. Machine-learning models are becoming indispensable components even in systems that require safety-critical environmental perception and decision-making, such as automated-driving systems [1]. For human society to accept safety-critical machine-learning systems however, it is important to develop common quality-assurance (QA) frameworks for managing the risks of using machine-learning models [3]. QA has an impact on the social receptivity, because it can be one of the approaches to deliver safety and security. In fact, recent accidents caused during the use of several experimental automated vehicles have revealed QA frameworks as imperative for addressing this upcoming social issue [2]. However, although the automated-driving technologies are being actively developed and proposed on a daily basis, QA concept and technologies to ensure safety and security have not been systematized yet. Therefore, in this study, we organize the review open QA problems on safety-critical machine-learning systems using in-vehicle automated-driving systems equipped with machine-learning models as an example. The contributions of this study are:

- Proposed points of view to clarify open QA problems on safety-critical machine-learning systems;
- Proposed possible open QA problems on safety-critical machine-learning systems along with the POV;
- Shown the recent trends addressing the open problems;
- Shown ideas to address some of the open problems;
- Shown the relevant industrial fields and the next research directions to address the open problems.

## 2 QUALITY ASSURANCE OF IN-VEHICLE AUTOMATED-DRIVING SYSTEMS

An automated-driving vehicle is a vehicle that operates without human input. Automated driving is not built as a stand-alone system in a vehicle. To realize automated driving, a system consisting of clouds, roadside devices (fog), and automated-driving vehicles (edge) [6] are necessary that creates and updates high-precision digital maps [5], and co-operates with peripheral vehicles. An in-vehicle automated-driving system installed in a vehicle is composed of multiple subsystems for perception, planning, and control, and it realizes automated-driving operations in cooperation with the clouds and roadside units. Each perception, planning, and control subsystem may contain machine-learning models, as necessary. Supervised-learning models [7] and decision-reinforcement learning models [8] can be used for perception and planning, and conventional control theory can be used for control, respectively. On the other hand, in order to perform QA on a system, it is necessary to predefine quality standards and a QA process in advance, and strictly follow them at development time. In this study, we identify open problems related to the three levels: the in-vehicle automated-driving system, the subsystems, and the machine learning model. Thus, we use an automated-driving system as an example of a safety-critical machine-learning system. Then, we investigate the open QA problems in terms of the three steps of the development process: design (requirements and specifications), verification (including validation), and operation (including maintenance). The three levels and three steps are shown in Fig. 1. Open problems that can be addressed using conventional methods are beyond the scope of this research study and are therefore excluded. Notably, many of the open problems considered in this paper do not occur only in automated-driving systems, but in safety-critical systems in general.

## 3 OPEN PROBLEMS OF IN-VEHICLE AUTOMATED-DRIVING SYSTEMS

In-vehicle automated-driving systems are automated-driving systems that are installed in vehicles. In the three layers of the system operating on an automated vehicle: the in-vehicle automated-driving system, the subsystems, and the machine learning model, only the former is exposed to the physical environment. Therefore, it is necessary to consider data on the physical environment more than theoretical assumptions when designing and verifying an in-vehicle automated-driving system. The open problems in this paragraph are to capture the operational physical environments of automated-driving systems, and address the intractableness of field operation testing (FOT).

**Requirement**: QA requires limitations, or warranty scopes. Therefore, a machine-learning system must be implemented in a predefined environment. In the automobile industry, this is called the operational design domain (ODD), and it essentially corresponds to the driving conditions with which the system is compatible [10]. Efficiently collecting driving scenarios on actual roads from past field operational tests and customer data remains a huge task and an open challenge in QA, but it is necessary for defining assumed environments and verifying the developed QA system against them. In addition, the evaluation criteria (such as test coverage) must be defined for the collected scenarios, in order to measure whether they are sufficient to ensure safety and security.

**Verification**: The simplest approach to guaranteeing the quality of an in-vehicle automated driving system is through verification against actual data. Accumulating a large number of safe automated-driving trips, as long distance as with human drivers, will effectively demonstrate that automated-driving systems are as safe as human drivers. However, obtaining statistically significant results would require conducting FOT by driving for hundreds of thousands to hundreds of billions of miles FOT [11]. In order to verify the system within a realistic time-frame, it is necessary to accelerate the experiment by simulation, reproducing the corner-case scenarios on test courses with a short mileage, *i.e.*, scenarios with extremely low probability of occurrence and difficult to statistically obtain (rare) by FOT on the actual road.

### 3.1 Trend

The German PEGASUS project is a joint initiative of vehicle manufacturers (OEM), suppliers, tool vendors, certification organizations, and research institutes, which aims to define standard QA methods for automated-driving systems [12]. The purpose of this project is to clarify the expected performance level and evaluation criteria of automated-driving systems through scenario-based verification. The scope of the project includes standard test procedures, continuous and flexible tool chains, integration of tests into the development processes, cross-company test methods, requirements definition methods, driving scenarios, a common database of scenarios, as well as creating safety statements based on them. Each scenario is defined by three levels of abstraction: the functional scenario (natural language description), the logical scenario (parameterized model), and the concrete scenario (model with specified parameters). First, in the concept phase, natural languages are used to describe the functional scenario without ambiguous terms, in a way that can be understood by human experts. Next, in the system development phase, we define the parameters and their distributions, and describe the logical scenario in a specified data format. In the test & verification phase, specific parameters are sampled from the parameter distributions, and the scenario is detailed into a concrete scenario, so that it can to be executed (simulated) with test tools. Scenarios are collected from test drives and the market to demonstrate that systems are equal to or better than human drivers. Regular scenarios are continuously tested by simulation, and those critical are tested through artificially configured environments on test courses. The PEGASUS project is an excellent example of continuous definition of requirements for (in-vehicle) automated-driving systems and their verification. The open problems in QA for the in-vehicle automated-driving system level require statistics and simulation techniques to develop solutions for scenario collection and verification, respectively. Thereafter, it is expected that the automobile industry (especially vehicle manufacturers) will select the appropriate solutions and lead the standardization process.

## 4 OPEN PROBLEMS OF SUBSYSTEMS

Subsystems run on top of the in-vehicle automated-driving system to provide special functionalities. The open problems in this paragraph are to design subsystems which include machine learning models, by considering and applying the characteristics of machine learning, such as uncertainty and "Change Anything Change Everything" (CACE) [15] .

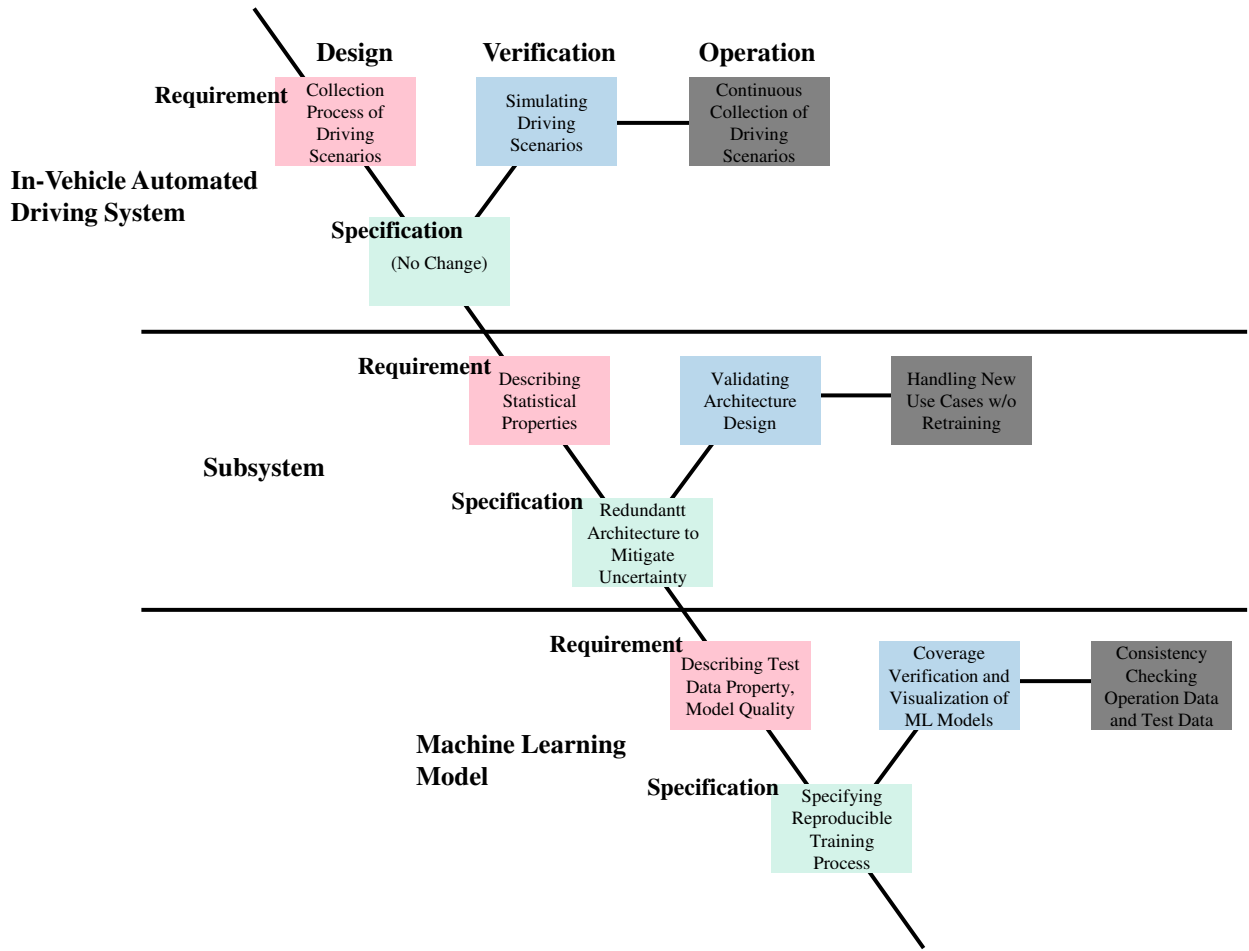

**Figure 1: Development Process of Safety-Critical ML Systems (Example: In-Vehicle Automated-Driving Systems)**

**Requirement**: Subsystems with machine-learning models have uncertainty derived from the machine-learning models. Therefore, it is necessary to develop methods for describing this uncertainty as a statistical measure.

**Design**: Thus, QA must be conducted independently for each subsystem including those incorporating machine-learning models. As a result, QA of subsystems that do not use machine-learning models can be performed through conventional methods, with minimal change in the development process. Subsystems with different functionalities, such as perception and planning, may require different QA approaches. For example, the same QA approach cannot be applied to perception with a Convolutional Neural Network [7] and to planning with a Deep Q-Network [8]. In general, a machine-learning model cannot achieve 100% accuracy on test data [13], and as accuracy approaches 100%, further performance improvement becomes difficult. Therefore, optimizing machine-learning models are not the only way to improve subsystem performance. In fact, a subsystem should be subdivided and verified based on its development process, combining the conventional deductive approach with the inductive approach that is unique to machine learning [14]. Rigorous breakdown of subsystem requirements into machine-learning model requirements is essential for safety-critical machine-learning systems. In this process, safety analysis methods and processes are important, such as: black box verification; encapsulation of machine-learning models by rule-based safeguarding; redundant & diverse architecture that mitigates and absorbs the uncertainty of machine learning models; system modification patterns to cope with additional requirements without modification of the machine-learning model, in order to avoid the principle of CACE as much as possible.

**Verification**: Moreover, in order to prevent reworking after training the machine-learning models, which is a heavy heuristic process, it is necessary to verify the validity of the redundant architecture. Assuming that a machine learning model behaves probabilistically based on the predetermined uncertainty, we would like to know if the uncertainty of the entire system is absorbed or mitigated, before training the model.

## 4.1 Trend

System-level technologies STAMP/STPA [16] and GSN [17] have received attention in recent years. They are highly promising, but they will need an extension that can express the uncertainty related to the machine-learning models. In addition, SOTIF, which is a safety standard/process concerning performance limits of normal functions, is also focused on securing functions with uncertainty even under normal conditions [18]. We consider that the most relevant technical area for QA at the subsystems level is software engineering. Subsystems are the intermediate layer between the deductive (manual) development of systems and the inductive (data driven) development of machine-learning models. The safety-critical properties of systems should be managed at the subsystem level and handed to the machine-learning models.

## 5 OPEN PROBLEMS OF MACHINE LEARNING MODELS

A machine-learning model is acquired by executing a training algorithm (training) with a model structure and training data sets as inputs. A machine-learning model before determining the parameters by training is called a model structure in this paper. Trained machine models are evaluated using test datasets [19]. The open problems in this paragraph are to deductively design the requirements of the machine learning models and their test data, as the final step manual engineering in machine-learning systems.

**Requirement**: In machine learning, we consider that the roles of training data and test data are different. While the training data is used for improving the performance of the machine-learning model, the test data should be used to accurately reflect the environmental conditions in operation. For example, when all the data obtained at the time of development are randomly divided, some of them are used as training data and the others are test data. In this way, the training data and the test data are approximately equally distributed, but the relationship to the operational data is unknown. Particularly in a safety-critical machine-learning system, it is necessary to design the test data distribution considering the operational environment, in which the system will actually be operated in the future, and to collect test data based on the designed data distribution. In this way, by accepting an a-priori viewpoint of the test data distribution, we can deductively define the assumed environment, and inductively collect data based on the requirements of the overlaid subsystem. Moreover, by assuming the distribution of the test data, we can now discuss their distribution, and define the operation domain with concrete design information.

Even in the current development of in-vehicle automated-driving systems, the test data would be collected assuming the operational environment, in order to make the distribution at the time of operation and the distribution of the test data as consistent as possible. However, the methods of describing the assumed environment of machine-learning models are not organized. In particular, we need methods to describe the properties of the test data. Furthermore, since the required performance may change depending on the assumed environment, it is necessary to express the association between the assumed environment and the required performance. Test data should have attributes such as time and weather, and their distributions, which are based on the assumed environment. Each condition of the test data distribution should have a confusion matrix which machine learning models will have them as desired value (Table 1).

In Table 1, the confusion matrix (accuracy) is shown as a requirement of the machine-learning model, but there are other quality characteristics of machine-learning models that should be investigated. In order to use machine-learning models in safety-critical systems, it is necessary to define and evaluate models by the performance indicators from various perspectives besides accuracy. If we refer to the software quality model [20], then accuracy (confusion matrix), resource behavior (required memory), time behavior (execution time), fault tolerance (robustness against Adversarial Examples [22], hereinafter referred to as AE), analyzability (ease of internal visualization or interpretability), and testability (simple model structure advantageous for formal verification) appear to be relevant perspective (performance indicator) to machine-learning models.

**Design**: A machine-learning model is automatically obtained by training the model structure using the training data. Thus, specifications cannot be designed a priori. This limitation is essential, because high-performance machine-learning models are realized by learning high-dimensional parameters from data that engineers cannot manually specify. However, in the development of a safety-critical machine-learning system, at a minimum, it is necessary to record the model structure, the training data, and the training system (including hyper parameters, initial parameters, random number seeds, etc.) to secure the reproducibility of the training process. In this way, specification is achieved indirectly.

Operational data tend to change with time, creating a major problem for the operation and maintenance of machine-learning models [21]. In this way, the deviation between test data used during development and operation data becomes large as time elapses from completion of development. It is important to check the consistency between the operational data and the test data (originally assumed environment), and to make the machine-learning models follow the operational data in a continuous maintenance process.

Table 1: top - sample environment (data distribution), bottom - sample requirements (confusion matrix)

|         |      | time |       |
|---------|------|------|-------|
|         |      | day  | night |
| weather | fine | 40% † | 30%  |
|         | rain | 20%  | 10% ‡ |

| fine × day † |            | prediction |         |
|--------------|------------|------------|---------|
|              |            | pedestrian | vehicle |
| actual       | pedestrian | 90%        | 10%     |
|              | vehicle    | 20%        | 80%     |

| rain × night ‡ |            | prediction |         |
|----------------|------------|------------|---------|
|                |            | pedestrian | vehicle |
| actual         | pedestrian | 85%        | 15%     |
|                | vehicle    | 15%        | 85%     |

† e.g., when fine day there are many pedestrians, therefore precision on pedestrian is prioritized.
‡ e.g., when rainy night there are many vehicles, therefore precision on vehicle is prioritized.

**Verification**: Some quality characteristics cannot be evaluated with test data. For example, when evaluating robustness against AE as fault tolerance, it is necessary to artificially generate perturbations around the original test data points. We can generate AE near each individual test data point and quantify robustness by maximizing the AE to which the model can give correct answers. Besides robustness, there are also demands for measurement methods of various quality characteristics as listed above.

## 5.1 Trend

Increasing stability against disturbance, that is, enhancing robustness is a key to QA. AE occurs when the image recognition model incorrectly recognizes slight noise that cannot be recognized by humans with high confidence [22]. AE is known to have model-independent versatility and is an issue that threatens the safety of automated-driving systems depending on image-recognition technology. Study on AE is important not only for defending intentional attacks such as false traffic signs, but also for understanding the formation of decision boundaries of machine learning model and realizing highly robust recognition functions.

Inference processes of advanced machine learning models such as neural networks (NN) are considered black boxes. In particular, safety-critical systems should exhibit interpretability and transparency. In this context, a black box refers to a situation, where, although feature activations can be observed physically, the actual phenomenon cannot be understood. Regarding interpretability, NN visualization [23] shows great promise. In NN visualization, one method intentionally performs occlusion on input data, and specifies the region where the inference result changes drastically as a region of interest; another method back-propagates activation values from the influencer nodes at the later feature extraction process to identify the region of interest. NN visualization may have use in cases of debugging during training, and validation of the training result (understanding the internal behavior of the trained NN). Therefore, one open problem is obtaining concrete use cases to address the interpretability and transparency of safety-critical machine-learning systems, by using NN visualization, interpretation, transparency methods.

In the previous paragraphs, we presented the current trends in machine-learning technology related to QA of machine-learning models. Furthermore, even in the field of theoretical computer science, toward a verification of machine-learning model, automatic coverage verification based on formal verification technology is becoming possible (Table 2). The technical areas necessary for QA at the machine-learning model level are machine learning and theoretical computer science.

## 6 CONCLUSION

With the development of rapid technology in recent years, machine learning has been used in various systems. In order to use machine learning in a safety-critical system such as an automated-driving system, it is necessary to demonstrate the safety and security of the system to society through QA. In this paper, taking automated driving as an example, we presented the open problems and corresponding research/industry trends from the viewpoint of design, verification, and operation for in-vehicle automated-driving systems, subsystems, and machine-learning models. As a result, we found that the automobile industry (standardization), statistics, software engineering, machine learning, and theoretical computer science are the relevant industrial fields and the next research directions to investigate for solving the open problems of in-vehicle automated-driving systems, subsystems, and machine-learning models, respectively.

Table 2: Trend on Automatic Coverage Verification for Machine-Learning Models

| | | What to evaluate | Target | Reference |
|---|---|---|---|---|
| Decision problem | Verification | **Global safety**: for any inputs, the outputs of MLP is within a bound. **Local safety**: Consistency between target and training data. Specifically, for training data(x',y'), if the input of MLP is near x', the output of MLP is near y'. | Multi Layer Perceptron | -Pulina & Tacchella. Challenging SMT Solvers to Verify Neural Networks. In AI Comm. -Pulina & Tacchella. An Abstraction-Refinement Approach to Verification of Artificial Neural Networks. In CAV'10. |
| | | **(ε,δ)-robustness**: if the difference of two inputs ≦ ε, then the difference of corresponding outputs < δ | DNN | Katz et al. Towards Proving the Adversarial Robustness of DNNs. In FVAV'17. |
| | | **(x,η,Δ)-Safe**: given a perturbation between layers (defined by function Δ) and a set of input η, the corresponding outputs of NN with perturbation and NN are same. | Feedforward multilayer NN | Huang et al. Safety Verification of Deep Neural Networks. In CAV'17. |
| | | Satisfiability of a given **first order prop. logic** with **ReLU constraints**. | DNN | Katz et al. Reluplex: An Efficient SMT Solver for Verifying Deep Neural Networks. In CAV'17. |
| | Falsification | **Signal Temporal Logic** (which describes car-level specifications) | In-vehicle system including ML component | Dreossi et al. Compositional Falsification of Cyber-Physical Systems with Machine Learning Components. In NFM'17. |
| | | **Whether CNN recognizes cars** | CNN (classification of cars) | Dreossi et al. Systematic testing of convolutional neural networks for autonomous driving. In RMLW'17. |
| Function problem | | **Adversarial frequency Φ(f,ε)**: Given a distribution of inputs, the probability of the outputs of classifier f are different for input i and inputs in its ε-neighbourhood. **Adversarial severity μ(f,ε)**: the expectation of ε with probability of input i where ε denotes the minimal value such that the outputs of i and its ε-neighbourhood are different. | NN (Image classification) | Bastani et al. Measuring Neural Net Robustness with Constraints. In NIPS'16. |
| | | **Maximum perturbation bound Φ_(m,k)**: Suppose the ANN outputs the confidences of labels. The maximum absolute value of perturbation of inputs, such that for given inputs with perturbation, the number of ANN output nodes are more than the output of the node for label M is ≦ k-1. | ANN (ReLU node, softmax output) | Cheng et al. Maximum Resilience of Artificial Neural Networks. In ATVA'17. |

In order to enable QA of safety-critical machine-learning systems, research on elemental technology is also necessary, but even if each company carries out its own QA based on its own concepts, the results cannot be widely accepted by human society. Experts from each specialized field as mentioned in this paper should gather together, create technically reasonable drafts, proposing QA processes, and form consensus (standardization) overviewing the worldwide trends of safety-critical machine-learning systems. To this end, the present study can form a basis for future discussion and interdisciplinary research, with the aim to create a draft QA process.